\input epsf.sty
\newdimen\psfigsize
\def\psfigure#1 #2 #3 #4 #5{
    \begin{figure}[tbh]
      \vbox{
        \null\vskip-0.1in\hskip#2
        \epsfxsize=#1
        \epsfbox{#4}
        \vskip -0.1in
        \caption {#5 \label{#3}}
        \vskip 0.0truein plus0.1truein
      }
    \end{figure}
}

\def\nabstar#1{\nabla\kern-0.5pt\smash{\raise 4.5pt\hbox{$\ast$}}
               \kern-4.5pt_{#1}}

\def\drvstar#1{\partial\kern-0.5pt\smash{\raise 4.5pt\hbox{$\ast$}}
               \kern-5.0pt_{#1}}

\def\newline{\relax\ifhmode\null\hfil\break\else\nonhmodeerr@\newline\fi}
\def\frac#1#2{{#1\over#2}}
\def\text#1{{\hbox{\rm #1}}}

\newcommand{\beq}{\begin{equation}}
\newcommand{\eeq}{\end{equation}}
\newcommand{\bea}{\begin{eqnarray}}
\newcommand{\eea}{\end{eqnarray}}
\def\Id{ \mbox{1\hspace{-1.2mm}I} }
\def\BE{\begin{equation}}
\def\EE{\end{equation}}
\def\BA{\begin{eqnarray}}
\def\EA{\end{eqnarray}}
\def\BAN{\begin{eqnarray*}}
\def\EAN{\end{eqnarray*}}

\def\tr{\mathrm{tr}}

\def\gm5{\gamma_5}

\def\BE{\begin{equation}}
\def\EE{\end{equation}}
\def\BA{\begin{eqnarray}}
\def\EA{\end{eqnarray}}
\def\BAN{\begin{eqnarray*}}
\def\EAN{\end{eqnarray*}}

\def\tr{\mbox{tr}}

\def\text#1{{\rm #1}}
\documentstyle[twoside,fleqn,espcrc2]{article}
\begin{document}
\title{Ginsparg-Wilson relation with $ R = ( a \gamma_5 D )^{2k} $ }
\author{Ting-Wai Chiu%
\address{Department of Physics, National Taiwan University,
Taipei, Taiwan 106, Republic of China.}
\thanks{%
This work was supported by the National Science Council, R.O.C. under the
grant numbers NSC89-2112-M002-017 and NSC89-2112-M002-079}}

\begin{abstract}

The Ginsparg-Wilson relation $ D \gamma_5 + \gamma_5 D = 2 a D R \gamma_5 D $
with $ R = ( a \gamma_5 D )^{2k} $ is discussed.
An explicit realization of $ D $ is constructed.
It is shown that this sequence of topologically-proper lattice
Dirac operators tend to a nonlocal operator in the limit
$ k \to \infty $. This suggests that the
locality of a lattice Dirac operator is irrelevant to its index.

\end{abstract}

\maketitle

\section{INTRODUCTION}

In general, the Ginsparg-Wilson relation \cite{gwr}
for a lattice Dirac operator $ D $ can be written as
\bea
\label{eq:ggw}
D \gamma_5 f(D) + g(D) \gamma_5 D = 0 \ ,
\eea
where $ f $ and $ g $ are analytic functions.
Then the fermionic action $ {\cal A}_f = \bar\psi D \psi $ is
invariant under the generalized chiral transformation
\bea
\label{eq:gct_1}
\psi     &\rightarrow& \exp[ \gamma_5 \theta f(D) ] \ \psi \ ,     \\
\label{eq:gct_2}
\bar\psi &\rightarrow& \bar\psi \ \exp[ \theta g(D) \gamma_5 ] \ ,
\eea
where $ \theta $ is a global parameter.
In particular, for $ f(D) = \Id - a R D $ and $ g(D) = \Id - a D R $,
(\ref{eq:ggw}) becomes
\bea
\label{eq:gwr_0}
D \gm5 + \gm5 D = a D ( R \gm5 + \gm5 R ) D \ ,
\eea
where $ R $ is any operator.
Since $ ( R \gm5 + \gm5 R ) $ commutes with $ \gm5 $, without loss,
we can set $ R $ to commute with $ \gm5 $,
\bea
\label{eq:R_5}
R \gm5 = \gm5 R \ .
\eea
Then (\ref{eq:gwr_0}) becomes the usual form of the GW relation
\bea
\label{eq:gwr}
D \gm5 + \gm5 D =  2 a D R \gm5 D \ .
\eea

In the continuum, the massless Dirac operator
$ {\cal D} = \gamma_\mu ( \partial_\mu + i A_\mu ) $ is chirally
symmetric and anti-hermitian, i.e.,
\bea
\label{eq:chiral_sym}
{\cal D} \gm5 + \gm5 {\cal D} = 0 \ , \\
\label{eq:anti-hermit}
{\cal D}^{\dagger} = -{\cal D } \ ,
\eea
which imply that $ {\cal D} $ is $\gamma_5$-hermitian,
i.e., $ {\cal D}^{\dagger} = \gm5 {\cal D} \gm5 $.

On the lattice, a lattice Dirac operator $ D $ cannot
satisfy (\ref{eq:chiral_sym}) or (\ref{eq:anti-hermit}) without
losing its other essential properties. However, we can still require
$ D $ to satisfy the $\gamma_5$-hermiticity,
\bea
\label{eq:hermit}
D^{\dagger} = \gm5 D \gm5 \ .
\eea
Multiplying $ \gm5 $ on both sides of (\ref{eq:gwr}), and using
(\ref{eq:hermit}), we obtain $ D^{\dagger} + D = 2 a D^{\dagger} R D $,
which gives $ R^{\dagger} = R $.

Thus we conclude that, in general, $ R $ is a Hermitian operator
commuting with $ \gm5 $. From (\ref{eq:gwr}),
the GW fermion propagator ( in the trivial sector ) satisfies
$ D^{-1} \gm5 + \gm5 D^{-1} = 2 a R \gm5 $,
which immediately suggests that $ R $ should be as local as
possible, i.e., $ R \simeq \Id $. Otherwise, it may cause additive
mass renormalization to $ D^{-1} $, which of course is undesirable.
In fact, it has been demonstrated that the most optimal $ R $ is
$ R = r \Id $ with some constant $ r $ \cite{twc99:10}.

Nevertheless, theoretically, it is rather interesting to consider $ R $ as
a function of $ D $, and to see how $ D $ behaves with respect to $ R $.
It turns out that, in general, the independent variable in the functional
form of $ R $ must be $ ( a \gm5 D )^2 $, if $ D $ is normal and
$\gm5$-hermitian. In particular, for $ R(t)=t^{n}, n = 0, 1, \cdots $,
the sequence of topologcially-proper $ D $ (\ref{eq:Dk}) tend to
a nonlocal Dirac operator in the limit $ n \to \infty $.
This suggests that the locality of a lattice Dirac is irrelevant
to its index.

First, recall that for any $ D $ which is normal
( $ D D^{\dagger} = D^{\dagger} D $ ) and $\gm5$-hermitian,
its eigensystem ( $ D \phi_s = \lambda_s \phi_s $ )
has the following properties \cite{twc98:4} : \\
{\bf (i)} eigenvalues are either real or come in complex conjugate pairs. \\
{\bf (ii)} the chirality ( $ \chi_s = \phi_s^{\dagger} \gm5 \phi_s $ )
           of any complex eigenmode is zero. \\
{\bf (iii)} each real eigenmode has a definite chirality. \\
{\bf (iv)} sum of the chirality of all real eigenmodes is zero
           ( chirality sum rule ). \\

Assume that the eigenvalues of $ D $ lying on a simple closed contour
in the complex plane. Then {\bf (i)} implies that the real eigenvalues
of $ D $ can only occur at two different points, say at zero and $ a^{-1} $.
Then the chirality sum rule reads
\bea
\sum_{\lambda_s = 0, a^{-1} } \phi_s^{\dagger} \gm5 \phi_s = 0 \ ,
\eea
which immediately gives
\bea
\label{eq:chi_sum_rule}
n_{+} - n_{-} + N_{+} - N_{-} = 0 \ .
\eea
where $ n_\pm $ denotes the number of zero modes of
$ \pm $ chirality, and $ N_{\pm} $ the number
of nonzero real ( $ a^{-1} $ ) eigenmodes of $ \pm $
chirality. Then we immediately see that any zero mode must be
accompanied by a real ( $ a^{-1} $ ) eigenmode with opposite chirality,
and the index of $ D $ is
\bea
\label{eq:index_D}
\mbox{index}(D) \equiv n_{-} - n_{+} = - ( N_{-} - N_{+} ) \ .
\eea
It should be emphasized that the chiral properties {\bf (ii), (iii)}
and the chirality sum rule (\ref{eq:chi_sum_rule})
hold for any normal $ D $ satisfying the $\gm5$-hermiticity,
as shown in Ref. \cite{twc98:4}. However, in nontrivial gauge
backgrounds, whether $ D $ possesses any zero modes or not
relies on the topological characteristics \cite{twc99:11} of $ D $,
which cannot be guaranteed by the conditions such as the locality,
free of species doublings, correct continuum behavior,
$\gm5$-hermiticity and the GW relation.

Therefore, we must require that $ D $ is normal and
$\gamma_5$-hermitian,
\BAN
D^{\dagger} D = D D^{\dagger}, \hspace{4mm}
D^{\dagger} = \gm5 D \gm5 \ .
\EAN
These two equations immediately give
\bea
\gm5 ( \gm5 D )^2 = ( \gm5 D )^2 \gm5  \ .
\eea
Thus we have found an example of $ R = ( a \gm5 D )^2 $ which
is Hermitian and commutes with $ \gm5 $. In general, $ R $ can be any
analytic function of $ ( a \gm5 D )^2 $, i.e.,
\bea
\label{eq:R_general}
R = f[ ( a \gm5 D )^2 ] \ , \hspace{2mm} \mbox{ $ f $ : analytic function.}
\eea
In particular, for $ f(x) = x^n $, it becomes
\bea
\label{eq:R}
R = ( a \gm5 D )^{2k} \ , \hspace{4mm} k = 0, 1, 2, \cdots
\eea
Substituting (\ref{eq:R}) into the GW relation (\ref{eq:gwr}), we obtain
\bea
\label{eq:gwr_f}
\gm5 D + D \gm5 = 2 a D ( a \gm5 D )^{2k} \gm5 D \ ,
\eea
which is equivalent to Fujikawa's proposal \cite{fuji00:4}
\bea
\label{eq:gwr_of}
\gm5 ( \gm5 D ) + ( \gm5 D ) \gm5 =  2 a^{2k+1} ( \gm5 D )^{2k+2} \ .
\eea

\section{A CONSTRUCTION OF $ D $ }

So far, the only viable way to construct a topologically-proper lattice
Dirac operator is the Overlap \cite{rn95}
\bea
\label{eq:overlap}
D = \frac{1}{2a} ( \Id + \gamma_5 \epsilon ), \hspace{4mm} \epsilon^2 = \Id
\eea
which satisfies the GW relation (\ref{eq:gwr}) with $ R = 1 $.
Since $ D $ (\ref{eq:overlap}) is normal and $ \gm5 $-hermitian, its
eigenmodes satisfy the chiral properties {\bf (i)-(iv)}.
There are many different ways to implement the Hermitian
$ \epsilon $ in (\ref{eq:overlap}).
However, it is required to be able to capture the topology of the gauge
background. That means, one-half of the difference of the numbers
( $ h_\pm $ ) of positive ( $ +1 $ ) and negative ( $ -1 $ )
eigenvalues of $ \epsilon $ is equal to the background
topological charge $ Q $,
\BAN
\label{eq:index_Q}
   \sum_x \tr ( a \gm5 D(x,x) )
 = \frac{1}{2} \sum_x \tr( \epsilon )
 = \frac{1}{2} ( h_{+} - h_{-} ) = Q
\EAN
where $ \tr $ denotes the trace over the Dirac and color space.
Otherwise, the axial anomaly of $ D $ cannot agree with the
topological charge density in a nontrivial gauge background.

An explicit realization of $ \epsilon $ in
(\ref{eq:overlap}) is the Neuberger-Dirac operator \cite{hn97:7}
with
\bea
\label{eq:eh}
\epsilon = \frac{ H_w } { \sqrt{ H_w^2 } }
\eea
where
\bea
\label{eq:Hw}
H_w = \gm5 ( \gamma_\mu t_\mu + W - m_0 a^{-1} ), \  0 < m_0 < 2
\eea
\bea
\label{eq:tmu}
t_\mu (x,y) = \frac{1}{2a} \ [   U_{\mu}(x) \delta_{x+\hat\mu,y}
                       - U_{\mu}^{\dagger}(y) \delta_{x-\hat\mu,y} ]
\eea
\BAN
W(x,y) = \frac{1}{2a} \sum_\mu [ 2 \delta_{x,y}
                     - U_{\mu}(x) \delta_{x+\hat\mu,y}
                     - U_{\mu}^{\dagger}(y) \delta_{x-\hat\mu,y} ]
\EAN

To generalize this construction for $ k > 0 $,
we multiply both sides of the GW relation (\ref{eq:gwr})
by $ R $ and redefine $ D' = R D $, then we have
\bea
\label{eq:gwr_1}
D' \gm5 + \gm5 D' = 2 D' \gm5  D' \ .
\eea
It can be shown that $ D' $ is $\gm5$-hermitian.
Now (\ref{eq:gwr_1}) is in the same form of the GW relation with $ R = 1 $.
Thus, one can construct $ D' $ in the same way as the Overlap
\bea
\label{eq:D'}
D' = R D = \frac{1}{2a} ( \Id + \gm5 \epsilon ), \hspace{4mm}
\epsilon^2 = \Id \ ,
\eea
provided that a proper realization of $ \epsilon $ can be obtained.
Using (\ref{eq:R_5}), we obtain
\bea
D' = R D = a^{-1} \gm5 ( a \gm5 D )^{2k+1} \ ,
\eea
which immediately yields
\bea
\label{eq:DD'}
D = a^{-1} \gm5 ( a \gm5 D')^{1/(2k+1)} \ ,
\eea
where the $(2k+1)$-th real root of the
Hermitian operator $ ( a \gm5 D ' ) $ is assumed.
Then (\ref{eq:DD'}) suggests that if the $ \epsilon $ in (\ref{eq:D'})
is expressed in terms of a Hermitian operator $ H $, i.e.,
$ \epsilon = \frac{H}{\sqrt{H^2}} $,
then $ H $ is required to behave as
$ ( \gamma_\mu {\cal D}_\mu )^{2k+1} $
in the continuum limit such that $ D $ can become
$ \gamma_\mu {\cal D}_\mu $ after
taking the $ ( 2k+1 ) $-th root in (\ref{eq:DD'}).
Thus, $ H $ must contain the term $ ( \gamma_\mu t_\mu )^{2k+1} $,
where $ \gamma_\mu t_\mu $ is the naive lattice fermion operator
defined in (\ref{eq:tmu}). Then another term must be added
in order to remove the species doublers in
$ ( \gamma_\mu t_\mu )^{2k+1} $. So, we add the Wilson term
to the $(2k+1)$-th power, i.e., $ W^{2k+1} $.
Finally, a negative mass term $ -\frac{1}{2} a^{-(2k+1)} $
%
%
is inserted such that $ \epsilon $
is able to detect the topological charge $ Q $ of the
gauge background, i.e.,
\bea
\label{eq:HQ}
\frac{1}{2} \sum_x \tr \left( \frac{H}{\sqrt{H^2}} \right) = Q \ .
\eea
Putting all these terms together, we have
$$
H = \gm5 \left[ (-1)^{k} (\gamma_\mu t_\mu)^{2k+1} + W^{2k+1} -
           \frac{a^{-(2k+1)}}{2} \right]
$$
which, at $ k=0 $, reduces to the $ H_w $ in Eq. (\ref{eq:Hw}).
Then (\ref{eq:DD'}) can be rewritten as
\bea
\label{eq:Dk}
D = a^{-1} \left( \frac{1}{2} \right)^{\frac{1}{(2k+1)}}
    \gm5 \left( \gm5 + \frac{H}{\sqrt{H^2}} \right )^{\frac{1}{(2k+1)}}
\eea
which agrees with Fujikawa's construction \cite{fuji00:4}.
Note that (\ref{eq:Dk}) can also be written in the form
$ D = D_c ( \Id + a R D_c )^{-1} $, where $ D_c $ is dependent on the order
$ k $. The general properties of $ D $ have been derived in \cite{twc00:5}.
In particular, the index of $ D $ is independent of the order $ k $,
and the eigenvalues of $ D $ fall on a closed contour
with their real parts bounded between zero and $ a^{-1} $. In Fig. 1,
the eigenvalues of $ D $ are plotted for $ k = 0,1,2 $ respectively,
with the same background gauge field.

\section{IN THE LIMIT $ k \to \infty $}

One of the salient features of the sequence of topologically-proper
Dirac operators (\ref{eq:Dk}) is that the amount of
chiral symmetry breaking ( i.e., r.h.s. of (\ref{eq:gwr}) )
decreases as the order $ k $ increases.
However, at finite lattice spacing, the chiral symmetry breaking of $ D $
{\it cannot} be zero even in the limit $ k \to \infty $,
since $ D $ satisfies the GW relation (\ref{eq:gwr}).
In the limit $ k \to \infty $, the only possibility for $ D $ to break
the chiral symmetry is that $ D (p) $ becomes a piecewise
continuous function in the Brillouin zone, with discontinuities somewhere
at $ | p | \simeq a^{-1} $, and $ \gm5 D(p) + D(p) \gm5 = 0 $ for
$ | p | < a^{-1} $. Since such a $ D(p) $ is non-analytic at infinite
number of $ p $, the corresponding $ D(x,y) $ must be nonlocal in the
position space.

\psfigure 3.0in -0.2in {fig:lambda} {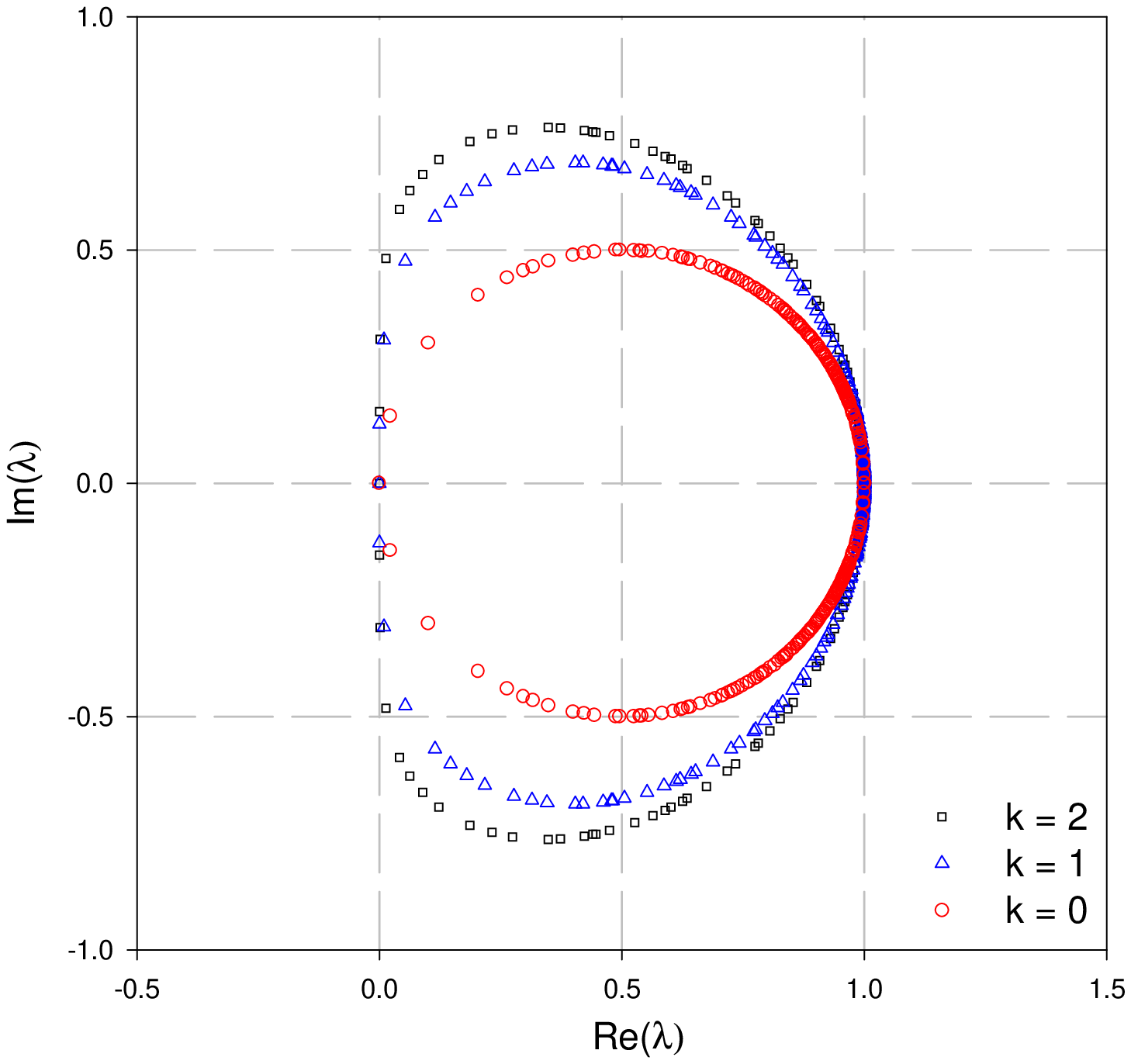} {
The eigenvalues of $ D $ in a nontrivial gauge background
for $ k=0,1,2 $ respectively. }

If $ D(x,y) $ is nonlocal in the free fermion limit
( i.e., $ U_\mu(x) \to 1 $ ), then it must be nonlocal
in any smooth gauge background. So, it suffices to examine
the non-analyticity of $ D(p) $ in the free fermion limit.

In the free fermion limit, the GW Dirac operator (\ref{eq:Dk}) in
momentum space can be written as
\bea
\label{eq:Dkp}
D(p)   =   D_0(p) + i \gamma_\mu D_\mu(p) \ ,
\eea
where
\BAN
D_0(p) = a^{-1} \left[ \frac{1}{2} \left( 1 + \frac{u(p)}{N(p)} \right)
                \right]^{\frac{k+1}{2k+1}} \ ,
\EAN
\BAN
D_\mu(p) = D_0(p)
\sqrt{\frac{1}{a^2 t^2(p)} \left( \frac{N(p)-u(p)}{N(p)+u(p)} \right) }
               \sin( p_\mu a ) \ ,
\EAN
\BAN
u(p) = a^{-(2k+1)}
         \left( \left[ \sum_\mu \ ( 1 - \cos( p_\mu a ) ) \right]^{2k+1}
                             - \frac{1}{2} \right)
\EAN
\BAN
t^2(p) = a^{-2} \sum_\mu \sin^2( p_\mu a ) \ ,
\EAN
\BAN
N(p) = \sqrt{ [t^2(p)]^{2k+1}+ [u(p)]^2 } \ .
\EAN

At the zeroth order ( $ k = 0 $ ), $ D(p) $ is analytic in the Brillouin zone.
However, in the limit $ k \to \infty $, $ D(p) $ tends to a nonanalytic
function with discontinuities at each point on two concentric hypersurfaces
inside the Brillouin zone \cite{twc00:8}. In four dimensions,
the inscribed hypersurface is specified by the equation
\BAN
\sum_{\mu=1}^{4} s_\mu^2 = 1 \ ,
\EAN
while the circumscribing one by
\BAN
6 + \sum_{\mu = 1}^{4} c_\mu ( c_\mu  - 4 )
                 + \sum_{\mu=1}^3 \sum_{\nu > \mu}^{4} c_\mu c_\nu  = 0 \ ,
\EAN
where $ c_\mu = \cos( p_\mu a ) $ and $ s_\mu = \sin( p_\mu a ) $.
This implies that $ D(x,y) $ is nonlocal in the limit $ k \to \infty $.

On the other hand, $ D(x,y) $ is local at the zeroth order ( $ k = 0 $ ),
for gauge backgrounds which are sufficiently smooth at the scale of the
lattice spacing \cite{ml98:8a,hn99:11}.
Since the index of $ D $ (\ref{eq:Dk}) is independent of the order $ k $,
this suggests that the locality of a lattice Dirac operator is
irrelevant to its index. In other words, if a GW Dirac operator which
is $\gm5$-hermitian, doubler-free, and has correct continuum behavior
in the free fermion limit, {\it but} always has zero index
\cite{twc99:11}, then the cause may {\it not} be due to its nonlocality.

To summarize, we have provided a nontrivial example of $ D $ (\ref{eq:Dk}),
which is {\it topologically-proper} but tends to a {\it nonlocal} operator
in the limit $ k \to \infty $. Unlike the nonlocal and chirally-symmetric
Dirac operator
\bea
\label{eq:Dc}
D_c = a^{-1} \frac{ \sqrt{ H_w^2} + \gm5 H_w }
                  { \sqrt{ H_w^2} - \gm5 H_w }
\eea
which has poles ( i.e., the nonzero real eigenmodes )
for nontrivial background gauge fields
( as a consequence of the chirality sum rule \cite{twc98:4} ),
the present example (\ref{eq:Dk}) breaks the chiral symmetry according
to the GW relation, thus its nonzero real ( $ a^{-1} $ ) eigenmodes
are well-defined for any order $ k $.

\end{document}